\documentclass[preprint,aps,prd,nofootinbib,superscriptaddress]{revtex4}
\usepackage{hyperref}
\usepackage{graphicx}
\usepackage{slashed}
\usepackage{amsmath,amssymb}
\usepackage{hhline}
\usepackage{subfigure}
\usepackage{epstopdf}

\newcommand{\beq}{\begin{equation}}
\newcommand{\eeq}{\end{equation}}
\newcommand{\beqa}{\begin{eqnarray}}
\newcommand{\eeqa}{\end{eqnarray}}

\newcommand{\met}{\ensuremath{{\not\mathrel{E}}_T}}

\newcommand{\Bbar}{\,\overline{\!B}{}}
\newcommand{\Dbar}{\,\overline{\!D}{}}
\newcommand{\Kbar}{\,\overline{\!K}{}}
\def\B0bar{\Bbar{}^0}
\def\D0bar{\Dbar{}^0}
\def\K0bar{\Kbar{}^0}

\begin{document}

\title{\boldmath Searching for dark matter at LHC with Mono-Higgs production}

\def\wsu{Department of Physics and Astronomy, 
Wayne State University, Detroit, MI 48201, USA}
\def\mctp{Michigan Center for Theoretical Physics, 
University of Michigan, Ann Arbor, MI 48109, USA}
\def\ucsc{Santa Cruz Institute for Particle Physics and
University of California, Santa Cruz, CA 95064, USA}

\author{Alexey A.\ Petrov}
\affiliation{\wsu}\affiliation{\mctp}
\author{William Shepherd}
\affiliation{\ucsc}

\date{\today}


\begin{abstract}
We consider LHC searches for dark matter in the mono-Higgs channel using the tools of effective field theory. 
This channel takes unique advantage of the presence of $SU(2)_L$ breaking in those operators to avoid the 
need for any initial-state radiation, usually necessary to tag the production of invisible particles.
We find that sensitivities to parameters describing dark matter 
interactions with standard model particles are comparable to those from monojet searches for a subset of the 
usually-considered operators, and we present for the first time bounds from collider searches on operators
which couple DM to only the Higgs field or its covariant derivatives.
\end{abstract}

\maketitle

\section{Introduction}\label{Intro}

Weakly-interacting massive particles (WIMPs), stable neutral states, which exist in many extensions of the standard model (SM),
provide a good solution to cosmological dark matter problem~\cite{Jungman,Hooper:2009zm}. The most widely-used models, which 
also provide solutions to well-known shortcomings of the standard model (SM), include supersymmetry, which naturally gives rise to a 
WIMP candidate, provided that R-parity is conserved. Other natural candidates include Kaluza-Klein excitations of the SM 
particles, where stability can be realized thanks to momentum conservation in compactified extra dimension. 
Correct identification of the nature of dark matter will lend support to one or another extension of the standard model.
It is thus not surprising that much of the recent efforts in both high energy and astrophysical experiments has been 
directed towards searches for those states. 

There are many different approaches to studies of dark matter. The most systematic one involves setting up a model of 
dark matter based on its chosen internal characteristics (such as spin) and internal symmetries that govern its
interactions with luminous matter. This ``top-down" approach gives a complete set of predictions for experimental 
observables, but gives results which are only very narrowly applicable, depending crucially on the assumptions
made regarding the dark matter and all other assumed physics beyond the standard model.

A slightly more general approach is to consider a theory which contains the minimal possible field content to generate the dark
matter phenomenology of interest by introducing a minimal amount of new particles and operators constrained by the requirement of 
renormalizability~\cite{Burgess:2000yq,Kim:2008pp,simplified}. This technique goes 
by the name of simplified model analyses. There are many different simplified models that arise in particular limits of a given 
complete theory such as the minimal supersymmetric standard model (MSSM), and, in principle, all of them would need to 
be considered to get an even partially complete understanding of the possible dark matter phenomena in the complete model.

This approach has been further generalized by attempting to make an absolutely minimal number of assumptions. This philosophy,
in line with the technique of effective field theory, considers all interactions of dark matter with the standard model that are permitted by 
a minimal set of assumptions. In recognition that the operators describing DM interactions could generically be introduced by some heavy 
particles that have been integrated out of the spectrum, operators of dimensions higher than four must be included as well.
Explorations of this approach have sparked new interest in the impact of colliders for dark matter physics, and have led to many interesting results 
comparing various types of experiments studying dark matter~\cite{Hooper:2009zm,morepeople}. There are definite concerns as to the applicability of these models to LHC (and other) physics~\cite{Buchmueller:2013dya,Profumo:2013hqa}

We will utilize this effective field theoretic approach and consider a possible signature of dark matter production which has not yet been 
explored at colliders~\cite{Goodman:2010ku,ExperimentalSearches}.
While the concept of utilizing the SM Higgs boson as an integral part of the physics of dark matter is by no means a new 
one~\cite{MarchRussell:2008yu,LopezHonorez:2012kv,Low:2011kp,Baek:2011aa,Mitsou:2013rwa,Chen:2013gya},
considering that connection explicitly at colliders has not extended much beyond considerations of light dark matter,
where the decays of the Higgs into dark matter give the Higgs a large invisible width which can be 
constrained~\cite{MarchRussell:2008yu,LopezHonorez:2012kv,Low:2011kp}
now that we are learning more about the specific properties of this boson. Our current goal is to consider
the possibility of Higgs production at the Large Hadron Collider (LHC) in association with a pair of dark matter candidate particles. 
To this end we will focus on possible interactions which couple to both the Higgs field and the dark matter, allowing for the signal to 
be produced at leading order in all couplings of the theory. As this is a signature which is already present in the SM due to the 
production of $Zh$ with the subsequent decay $Z\to\bar\nu\nu$, we are able to recast a study which is focused on SM Higgs 
properties \cite{thestudy} to consider this additional contribution.

In the next section we will discuss the theoretical framework applicable to this particular search. Following that,
in section \ref{recast}, we will present our recasting of the current searches for $Vh$ production to bound dark
matter-Higgs associated production. We conclude in section \ref{Conclusions} with discussion of future directions
for this signature-based dark matter search, both theoretical and experimental.

\section{Effective operators and LHC observables}\label{LHC}

Let us write a set of effective operators that can possibly generate our experimental signature. In what follows we shall 
consider all possible operators suppressed by at most three powers of the new physics scale and study their implications for experimental signals. 
Throughout, we label the DM field as $\chi$, and for concreteness we have assumed that DM is a fermion and a singlet of the SM gauge group.

Because we aren't working with a complete model of DM and its associated physics, and we've posited only the 
existence of the DM field, the only interactions which DM can have with SM fields are non-renormalizable. 
The lowest dimension at which the dark matter can interact with SM fields under these assumptions is five. 
These operators are
\beq\label{Dim5H}
{\cal L}_5 = \frac{2 C_1^{(5)}}{\Lambda} \left|H\right|^2 \overline \chi \chi +
\frac{2 C_2^{(5)}}{\Lambda} \left|H\right|^2 \overline \chi \gamma_5 \chi.
\eeq
Throughout, $C_i^{(n)}$ are the effective Wilson coefficients that characterize the strength of Higgs-DM interactions of dimension $n$
in the effective theory and $\Lambda$ characterizes the scale at which the EFT description breaks down. 
$H$ represents the Higgs doublet field, which in the unitary gauge takes the usual form 
\beq
H=\frac{1}{\sqrt{2}}
\left( 
\begin{array}{c}
0\\
v+h(x) 
\end{array} \right) .
\eeq
In terms of the physical Higgs field $h$, and considering only the terms quadratic in the physical field, this operator can be written as
\beq\label{Dim5}
{\cal L}_5 = \frac{C_1^{(5)}}{\Lambda} \ h^2  \overline \chi \chi +
\frac{C_2^{(5)}}{\Lambda} \ h^2  \overline \chi \gamma_5 \chi.
\eeq
We do not expect to have a strong bound on the scale $\Lambda$ from those operators, for two reasons. 
First, Higgs production by itself is relatively rare at the LHC, with subsequent DM-Higgs interactions giving an additional 
suppression. Second, since the Higgs boson is in the s-channel, it has to be significantly off-shell in 
this process. We note that the differences between the scalar and pseudoscalar couplings to the DM pair, 
while very significant for dynamics in the low-velocity regime, are negligible at the LHC where all particles are produced relativistically.

Next, there are two operators of dimension six,
\beq\label{Dim6H}
{\cal L}_6 = \frac{C_1^{(6)}}{\Lambda^2} H^\dagger \overleftrightarrow D_\mu H \ \overline \chi \gamma^\mu \chi +
\frac{C_2^{(6)}}{\Lambda^2} H^\dagger \overleftrightarrow D_\mu H \ \overline \chi \gamma^\mu \gamma_5 \chi,
\eeq
where we defined a covariant derivative 
$D_\mu = \partial_\mu - i(g/2) \sigma^a W^a_\mu - i (g^\prime/2) B_\mu$.
Once again, in the unitary gauge and in terms of the physical Higgs field $h$, the operators that could generate the 
mono-Higgs signature can be derived from Eq.~(\ref{Dim6H}) are
\beq\label{Dim6}
{\cal L}_6 = \frac{i C_1^{(6)} m_Z}{\Lambda^2} h Z_\mu \ \overline \chi \gamma^\mu \chi +
\frac{i C_2^{(6)} m_Z}{\Lambda^2} h Z_\mu \ \overline \chi \gamma^\mu \gamma_5 \chi,
\eeq
where we defined $Z_\mu = (g^2+g^{\prime 2})^{-1/2} (g W^3_\mu - g^\prime B_\mu)$ and employed the 
well-known relation $2 m_Z = v \sqrt{g^2+g^{\prime 2}}$. With an off-shell $Z$ boson in the $s$-channel
this operator gives rise to the desired experimental signature. Once again, the presense or absense of the
$\gamma_5$ in the DM bilinear does not appreciably impact the collider phenomenology predicted by the operator.
 
There are similar four operators of dimension seven that involve Higgs doublets and their
derivatives,
\beqa\label{Dim7H1}
{\cal L}_{7H} &=& \frac{C_1^{\prime (7)}}{\Lambda^3} \left(H^\dagger H\right)^2 \ \overline \chi \chi +
\frac{C_2^{\prime (7)}}{\Lambda^3} \left(H^\dagger H\right)^2  \ \overline \chi \gamma_5 \chi,
\nonumber \\
&+&  \frac{C_3^{\prime (7)}}{\Lambda^3} \left| D_\mu H\right|^2 \ \overline \chi \chi +
\frac{C_4^{\prime (7)}}{\Lambda^3} \left| D_\mu H\right|^2  \ \overline \chi \gamma_5 \chi .
\eeqa
The part of ${\cal L}_{7H}$ that generates the mono-Higgs signature at the LHC can be written as
\beqa\label{Dim7H}
{\cal L}_{7H} &=& \frac{3 C_1^{\prime (7)}}{2} \frac{v^2}{\Lambda^3} h^2 \ \overline \chi \chi +
\frac{3 C_2^{\prime (7)}}{2} \frac{v^2}{\Lambda^3} h^2  \ \overline \chi \gamma_5 \chi,
\nonumber \\
&+&  \frac{C_3^{\prime (7)}}{2 \Lambda^3} \left( \partial_\mu h\right)^2 \ \overline \chi \chi +
\frac{C_4^{\prime (7)}}{2 \Lambda^3} \left( \partial_\mu h\right)^2  \ \overline \chi \gamma_5 \chi .
\eeqa
We do not expect strong constraints on $\Lambda$ from those operators, as they simply represent higher-order 
$1/\Lambda$ corrections to the operators discussed above. We have listed them here for completeness, but shall 
not consider them further.

There are also four operators of dimension seven which describe coupling of dark matter to the SM fermions $f$,
\beqa\label{Dim7f1}
{\cal L}_{7F} &=& \frac{2 \sqrt{2} C_1^{(7)}}{\Lambda^3} \ y_d \ \overline Q_L H d_R  \ \overline \chi \chi +
\frac{2\sqrt{2} C_1^{(7)}}{\Lambda^3}  \ y_u \ \overline Q_L \widetilde H u_R  \ \overline \chi \chi 
\nonumber \\
&+& \frac{2 \sqrt{2} C_2^{(7)}}{\Lambda^3}\ y_d \  \overline Q_L H d_R \ \overline \chi \gamma^5 \chi +
 \frac{2\sqrt{2} C_2^{(7)}}{\Lambda^3} \ y_u \  \overline Q_L \widetilde H u_R  \ \overline \chi \gamma^5 \chi + h.c.
\eeqa
Here $\widetilde H=i\sigma_2 H^*$ is the usual charge-conjugated Higgs field and we scaled the Wilson coefficients to introduce 
Yukawa couplings $y_f$ for each fermion flavor $f=u,d$ of up ($u$) or down ($d$) type. $Q_L$ is a standard electroweak 
doublet of left-handed fermions. This form of operators is invariant under electroweak $SU(2)_L$ group and also 
naturally suppresses DM couplings to the light fermions, and is well motivated by the Minimal Flavor Violation 
paradigm~\cite{MinFlav}. We assume the couplings $C^{(7)}_i$ are flavor-blind, but permit them to be complex. In terms of the 
physical field $h$ the Eq.~(\ref{Dim7f1}) can be written as 
\beqa\label{Dim7f}
{\cal L}_{7F} &=& \frac{Re\left(C_{1}^{(7)}\right)}{\Lambda^3} \ y_f \left(\overline f  f\right) h  \left(\overline \chi \chi\right) +
 \frac{Im\left(C_{1}^{(7)}\right)}{\Lambda^3}  \ iy_f \left(\overline f  \gamma_5 f\right) h  \left( \overline \chi \chi\right) 
\nonumber \\
&+& \frac{Im\left(C_{2}^{(7)}\right)}{\Lambda^3} \ iy_f \left(\overline f  f\right) h  \left(\overline \chi \gamma_5 \chi\right) +
 \frac{Re\left(C_{2}^{(7)}\right)}{\Lambda^3}  \ y_f \left(\overline f  \gamma_5 f\right) h  \left( \overline \chi \gamma_5 \chi\right)
\eeqa
Note that these operators are identical to those which have traditionally been known as D1-D4 in the previous 
literature~\cite{Goodman:2010ku} on effective theories of DM scattering and production,
with the sole difference being that the implied Higgs vev has been replaced by the dynamical Higgs field in these operators.
This is another case where the scalar versus pseudoscalar nature of the couplings is not important to the collider phenomenology.
We expect the strongest constraints to come from this and the next set of operators, even though they are 
operators of relatively high dimension.

There are also four operators that are formally of dimension 8 that describe DM couplings to the 
gluons and the physical higgs~\cite{Petrov:2013vka}, 
\beqa\label{Dim7g}
{\cal L}_8 &=& \frac{C_1^{(8)}}{\Lambda^3M_{EW}} \left(\bar\chi\chi\right) h \ G^{a \mu\nu}G^a_{\mu\nu} +
\frac{C_2^{(8)}}{\Lambda^3M_{EW}} \left(\bar\chi\gamma^5\chi\right) h \ G^{a \mu\nu}G^a_{\mu\nu}
\nonumber \\
&+& \frac{C_3^{(8)}}{\Lambda^3M_{EW}} \left(\bar\chi\chi\right) h \ G^{a \mu\nu} \widetilde G^a_{\mu\nu} +
\frac{C_4^{(8)}}{\Lambda^3M_{EW}} \left(\bar\chi\gamma^5\chi\right) h \ G^{a \mu\nu} \widetilde G^a_{\mu\nu}
\eeqa
where we choose $M_{EW}=v$. Note that the presence of $M_{EW}$ here makes these operators equivalent in power counting of the new
physics scale to the dimension seven operators above. In fact, similarly to the operators in equation \ref{Dim7f}, these are equivalent
to the well-known operators D11-D14 of~\cite{Goodman:2010ku} with a Higgs vev replaced by the dynamical field.
Once again the parity structure of the operator is largely irrelevant for collider experiments.

It is important to note that these `dimension seven' operators mix with those in equation \ref{Dim7f} due to diagrams analogous to those responsible for the
gluon fusion production of the Higgs boson. As calculated in \cite{Petrov:2013vka,Haisch:2012kf}, any interaction of the form given in 
equation \ref{Dim7f} also gives rise at one loop to the corresponding interaction in Eq.~(\ref{Dim7g}), especially for the 
heavy fermions $f$. The importance of these higher-order operators coupling directly to gluons is also enhanced by the
large gluonic luminosity of the LHC.

\section{Mono-Higgs at LHC}\label{recast}

The characteristic signature of these interactions coupling dark matter to the Higgs boson is the production of a Higgs in association with missing energy. This, of course, suffers from a SM background from the associated production of a Z boson and a Higgs, with the Z subsequently decaying to neutrinos. However, this also presents an opportunity to immediately bound these interactions, as the ``background'' process has already been searched for by the LHC collaborations. In this section we recast, as well as possible, the results of the CMS search to apply to this new signal, with an assumed background of the SM signal strength.

The CMS search~\cite{thestudy} utilized multiple layers of boosted decision trees (BDTs) which we are not able to reproduce reliably. However, from their plots it is clear that all of their statistical power in differentiating signal from background comes from the highest of their three bins in missing energy. Thus, we construct our rudimentary comparison to the CMS analysis by requiring that events pass all of the cuts that were required to be used to train the BDTs in the high $\met$ region, and compare the accepted cross section of the SM $Zh$ signal to that predicted by each of the models we consider.

All of our samples are generated using MadGraph 5 \cite{Alwall:2011uj}, with parton showering and hadronization by PYTHIA \cite{Sjostrand:2006za} and rudimentary detector simulation 
using DELPHES \cite{delphes} tuned to emulate the CMS detector.  We generate 100000 events at the parton level for all signal samples to minimize the statistical errors induced 
by small acceptances. We have chosen as representative cases from each of those described above the unique operator which is parity-even in each bilinear. As discussed, choosing a different parity structure for the operator will not significantly affect the collider bounds that can be derived for that class of operator. We generate signal samples of the monohiggs final state, requiring the higgs to decay to a bottom quark pair.

The theoretical interpretation of the Wilson coefficients and the new physics scales are very different, but it is only the effective coupling which can be bounded under the assumption that the effective theory accurately describes the LHC physics. Thus, we have chosen to fix the Wilson coefficients and bound the new physics scale. We have chosen $C_i^{(n)}=1$ for most of our candidate interactions, with the only exception being the gluonic coupling, where we choose an interaction strength of $C_1^{(8)}=-\frac{g_s^2}{12\pi^2}\left(1+\frac{7m_h^2}{120m_t^2}+\frac{m_h^4}{168m_t^4}+\frac{13m_h^6}{16800m_t^6}\right)$, which is that induced by $C_1^{(7)}=1$ by the top loop in the large $m_t$ limit \cite{Rizzo:1979mf}. While this limit is not particularly well-justified, it was found by~\cite{Haisch:2012kf} that this amounts to at worst a factor of less than 2 overestimate of the bound strength on $\Lambda$ from monojet searches utilizing the same loop-level relationship between quark and lepton couplings.

The requirements we impose on an event to be accepted are:

\begin{itemize}
\item $\met > 170$ GeV
\item $\Delta\phi\left(\met,j\right)>0.5\forall j | P_T>25$ GeV, $|\eta|<2.5$
\item 0 leptons with $P_T>20$ GeV
\item 2 $b$-tagged jets, one with $P_T>60$, other with $P_T>30$
\item $M_{bb}<250$ GeV
\item $P_{T,bb}>130$ GeV
\item $\Delta\phi\left(\met,bb\right)>2$.
\end{itemize}

We note that b-tagging in particular is a place where our analysis and that of the CMS collaboration is likely to differ significantly. In particular, the CMS analysis placed differing requirements on the two $b$-tagged jets, requiring that one was very tightly identified as a $b$ quark and allowing the second to be less tightly identified. We lack the freedom to change $b$-tag criteria straightforwardly using the DELPHES package, and so simply require both jets to be tagged. It is reasonable to expect that the differences in $b$-tagging efficiencies will be independent of the underlying interaction which produces the Higgs boson that subsequently decays to the $b$ quarks.

Applying the above requirements to the tree-level SM signal we find a total rate for $pp\to Zh$ with $Z\to\bar\nu\nu$ and $h\to\bar bb$ of 677 fb, and an acceptance of $0.64\%$ for the cuts applied above, giving a total accepted cross section of 4.30 fb. This is somewhat higher than the rate which can be reconstructed by considering the histograms shown in the CMS analysis, but the difference is likely due to the difference in $b$-tagging requirements and should therefore be universal for signal and background. We thus adopt the value of 4.3 fb accepted cross section as our definition of the SM signal strength $\mu=1$.

\begin{figure}
\subfigure[\label{fig:direct} Bounds on $\Lambda$ for the operator coupling DM and $h$ directy to SM quarks.]{\includegraphics[width=0.48\linewidth]{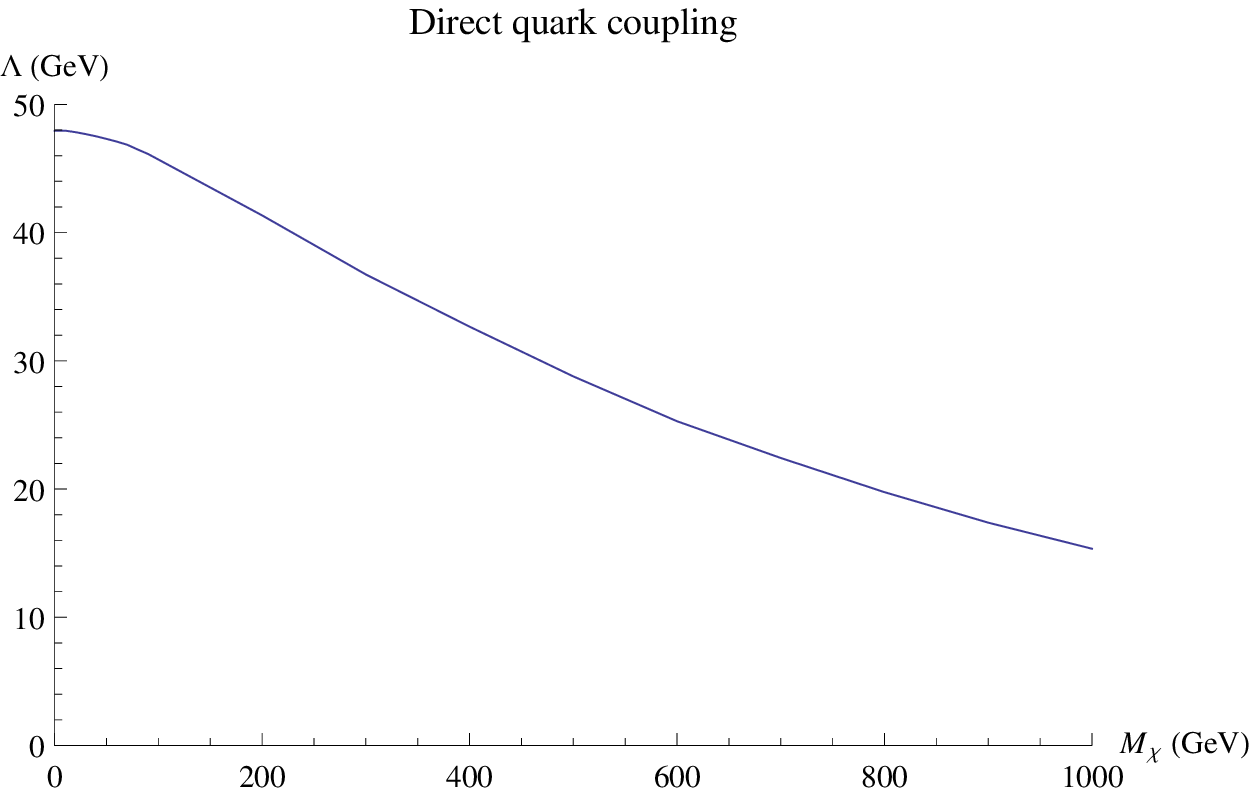}}
\subfigure[\label{fig:glue} Bounds on $\Lambda$ for the operator coupling DM and $h$ to gluons.]{\includegraphics[width=0.48\linewidth]{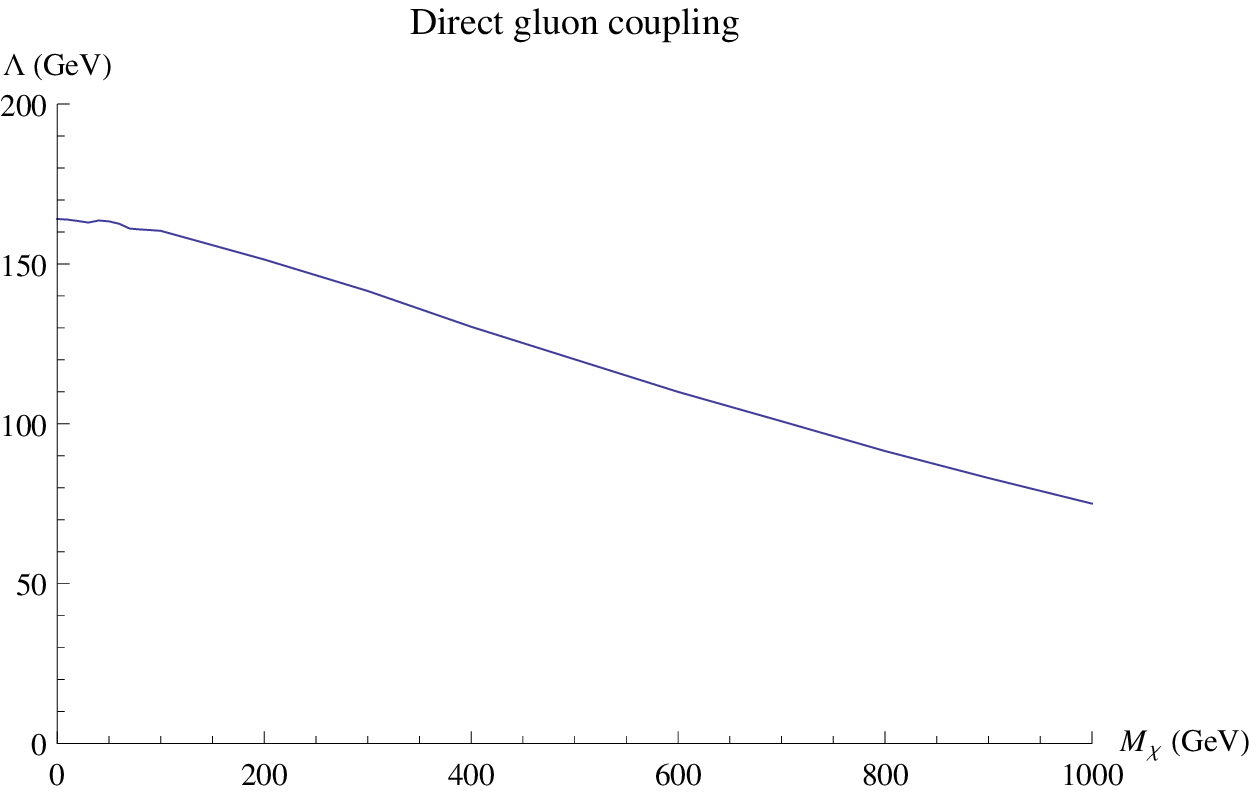}}
\\
\subfigure[\label{fig:higgs} Bounds on $\Lambda$ for the operator coupling DM directly to the SM Higgs.]{\includegraphics[width=0.48\linewidth]{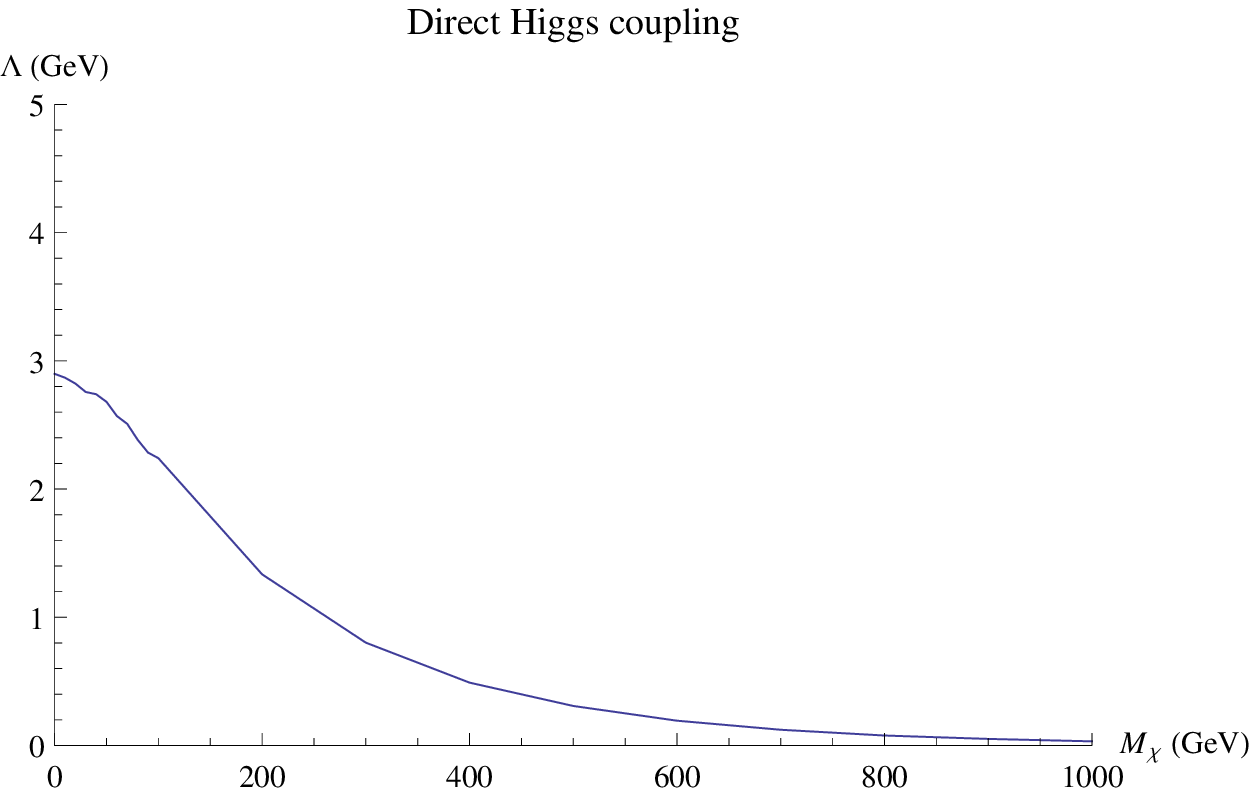}}
\subfigure[\label{fig:hz} Bounds on $\Lambda$ for the operator coupling DM to the Higgs and Z boson.]{\includegraphics[width=0.48\linewidth]{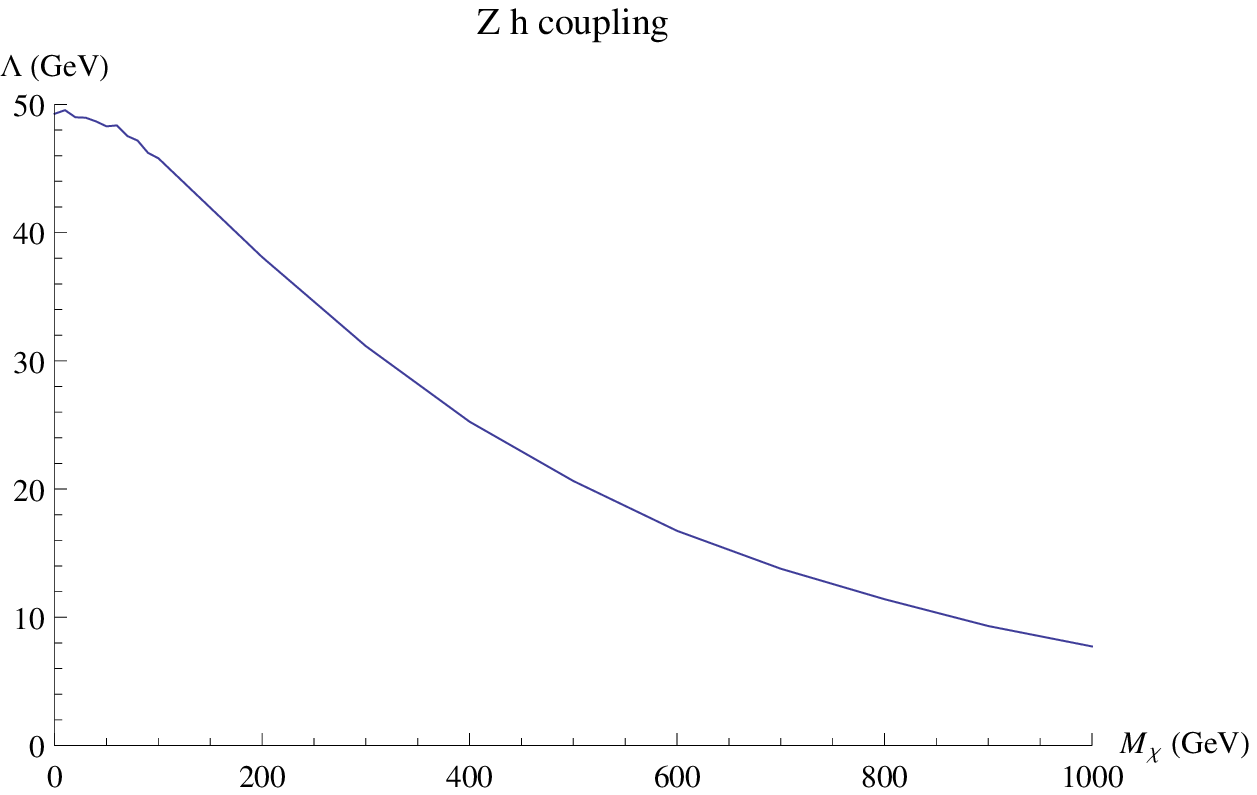}}
\caption{Bounds derived from the recast search for $h+\met$ on the suppression scale $\Lambda$ on each of the four considered operators.}
\label{fig:bounds}
\end{figure}

We then proceed to apply the above analysis cuts to each of the interactions presented in Sec. \ref{LHC} and find the suppression scale $\Lambda$ appropriate for each to give a contribution to the signal rate that saturates the $2\sigma$ bound quoted by CMS. The measured value of CMS is $\mu=1.04\pm0.77$, and we assume the presence of a SM Higgs signal (which we take to be very well modeled), so the $2\sigma$ bound on new physics contributions is $\mu=1.48$ or $\sigma*A=6.4$ fb. The resulting bounds on $\Lambda$ in each model are presented in Fig.~\ref{fig:bounds}, assuming that all appropriate Wilson coefficients are of order one. Alternatively,
one can fix a NP scale and put constraints on the Wilson coefficients of each operator. This procedure is completely equivalent to the one chosen in 
this paper, so we shall only present constraints on the new physics scale $\Lambda$, with the assumed Wilson coefficients discussed above.

We considered dark matter candidates with masses up to 1 TeV, resimulating at each point to capture any changes in analysis efficiency with the dark matter mass. All such deviations are relatively minor, with the dominant effect being due to the change in total production cross section for the mono-higgs final state.

\begin{figure}
\includegraphics[width=0.48\linewidth]{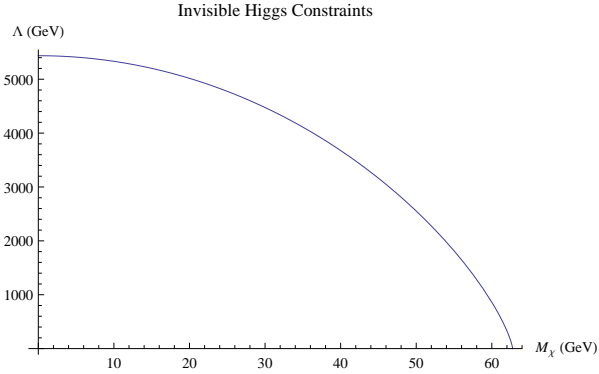}
\caption{\label{fig:invwidth}Bounds on the suppression scale of the operator coupling directly to the Higgs boson due to the known limits on the invisible width of the Higgs \cite{invwidth}.}
\end{figure}
As can be seen from Fig. \ref{fig:higgs}, the bounds on couplings of DM of the type introduced in Eq.~(\ref{Dim5}), i.e. directly to the SM Higgs, are very 
weak, but this is to be expected as it requires a far off-shell Higgs in the s-channel, as discussed in Sec.~\ref{LHC}. Stronger bounds on this operator can 
be obtained from the constraint on the Higgs invisible width for $m_h>2m_\chi$, and are presented in Fig. \ref{fig:invwidth}. Note that this figure uses a different scale for the dark matter mass axis from the others. Bounds on the 
coupling to quarks (Fig. \ref{fig:direct}) are comparable with those derived from other missing energy searches considering the similar operator with the 
Higgs vev inserted, and the bounds on couplings to gluons, shown in Fig. \ref{fig:glue}, while less powerful than those from monojets, are similar. These operators do not require any propagator to go off-shell, and as such inherit their only dependence on the dark matter mass from the PDFs. The bounds derived on the coupling of DM to 
the Higgs and Z boson (shown in Fig. \ref{fig:hz}) are stronger than we anticipated, as they suffer from the same off-shell suppression as in the case coupling 
directly to the Higgs pair, but the much stronger production of the Z boson versus the Higgs allows reasonable bounds to be derived nonetheless. In this case we can see the additional dependence on the dark matter mass due to the $s$-channel $Z$ boson being required to go further off-shell. Similar behavior can be seen in the direct higgs coupling case, but the bounds are so weak that the dependence is certainly irrelevant to their interpretation.

\section{Conclusions}\label{Conclusions}

We estimated the current bounds on dark matter interactions due to the possible associated production of dark matter pairs and a Higgs boson, 
assuming that the Higgs is SM-like apart from the introduced interaction with dark matter. We emphasize that two of the operators we have bounded are identical to those considered previously in the context of other collider dark matter searches, particularly the 
difficult to constrain D1 and the strongly constrained D11. In their previously-used form these operators had the Higgs vev 
explicitly introduced to give a functionally lower-dimensional operator, but the opportunity to not require relatively unlikely initial-state radiation is 
recovered by retaining the dynamical Higgs in the operators. The experimental channel with the signature described in the paper offers a new and 
competitive probe of those interactions. The two other operators we have considered have not been previously constrained by any collider search.

The bounds we find are somewhat weakened by the fact that the searches available for recasting to this new model have been constructed to have 
sensitivity to all possible $Vh$ leptonic final states and have been tailored specifically for the expectations derived from that signal model. A separate 
search which considered the signature of Higgs and missing energy more generally, or these models in particular, would likely give a significant 
improvement in the bounds available from this possible signature of dark matter production.

It is very interesting that the bounds are very different on the operators which couple to heavy quarks from those which couple to gluons, as we know that 
these two mix when higher orders in QCD are considered. An interesting theoretical undertaking relating to this signature would be to consider the loop 
amplitude which mixes the two within various simplified models which give rise to the operators considered here. One source of these couplings of dark 
matter to quarks and the Higgs is a squark, and thus the validity of loop calculations using only the effective operator is suspect as loop momentum 
should flow through the mediator as well as the particles external to the operator. We reserve this question for future study.

\section*{Acknowledgments}

We would like to thank Filip Moortgat for bringing to our attention study of $Zh$ production performed by CMS collaboration.
The work of AAP was supported in part by the U.S.\ Department of Energy under contract DE-FG02-12ER41825.
WS is partly supported by the U.S.\ Department of Energy under contract DE-FG02-04ER41268. 
The authors would like to thank Kavli Institute for Theoretical Physics at
the University of California Santa Barbara for hospitality where part of this work was performed and 
supported in part by the National Science Foundation under Grant No. NSF PHY11-25915.


\end{document}